\documentclass{article}

\usepackage{graphicx}
\usepackage{amssymb, amsmath, amsthm, epsfig, url}

\newtheorem{theorem}{Theorem}[section]
\newtheorem{lemma}[theorem]{Lemma}
\newtheorem{proposition}[theorem]{Proposition}
\newtheorem{definition}[theorem]{Definition}
\newtheorem{example}[theorem]{Example}

\usepackage[labelfont=bf]{caption} 
\newcommand{\bfa}{\mathbf{a}}
\newcommand{\bfb}{\mathbf{b}}
\newcommand{\bfe}{\mathbf{e}}
\newcommand{\bff}{\mathbf{f}}
\newcommand{\bfv}{\mathbf{v}}
\newcommand{\x}{\mathbf{x}}
\newcommand{\y}{\mathbf{y}}
\newcommand{\zero}{\mathbf{0}}
\newcommand{\zz}{{\mathbb Z}}
\newcommand{\cc}{{\mathbb C}}

\begin{document}


\title{Computing all Space Curve Solutions of Polynomial Systems
by Polyhedral Methods\thanks{This material is based upon work 
supported by the National Science Foundation under Grant No. 1440534.}}

\author{Nathan Bliss \and Jan Verschelde}

\date{University of Illinois at Chicago \\
Department of Mathematics, Statistics, and Computer Science \\
851 S. Morgan Street (m/c 249), Chicago, IL 60607-7045, USA \\
{\tt {\{nbliss2,janv\}@uic.edu}}}

\maketitle
\begin{abstract}
A polyhedral method to solve a system of polynomial equations
exploits its sparse structure via the Newton polytopes of the polynomials.
We propose a hybrid symbolic-numeric method to compute a Puiseux series
expansion for every space curve that is a solution of a polynomial system.
The focus of this paper concerns the difficult case when the leading
powers of the Puiseux series of the space curve are contained in the
relative interior of a higher dimensional cone of the tropical prevariety.
We show that this difficult case does not occur for polynomials
with generic coefficients.  To resolve this case, we propose to apply
polyhedral end games to recover tropisms hidden in the tropical prevariety.

\end{abstract}

\section{Introduction}
 
In this paper we consider the application of polyhedral methods
to compute series for all space curves defined by a polynomial system.
Polyhedral methods compute with the Newton polytopes of the system. 
The {\em Newton polytope} of a polynomial is defined as the convex hull
of the exponents of the monomials that appear with a nonzero coefficient.

If we start the development of the series where the space curve meets
the first coordinate plane, then we compute Puiseux series.
Collecting for each coordinate the leading exponents of a Puiseux series
gives what is called a {\em tropism}.
If we view a tropism as a normal vector to a hyperplane,
then we see that there are hyperplanes with this normal vector
that touch every Newton polytope of the system at an edge or 
at a higher dimensional face.
A vector normal to such a hyperplane is called a {\em pretropism}.
While every tropism is a pretropism, not every pretropism is a tropism.

In this paper we investigate the application of a polyhedral method
to compute all space curve solutions of a polynomial system.
The method starts from the collection of all pretropisms,
which are regarded as candidate tropisms.
For the method to work, we focus on the following questions.

\noindent {\bf Problem Statement.}
Given that only the space curves are of interest, can we ignore the
higher dimensional cones of pretropisms?
In particular, if some tropisms lie in the interior of higher dimensional cones
of pretropisms, is it then still possible to compute Puiseux series
solutions for all space curves?

\noindent {\bf Related Work.}
In symbolic computation, new elimination algorithms for sparse systems
with positive dimensional solution sets are described in~\cite{HJS14}.
Tropical resultants are computed in~\cite{JY13}.
Related polyhedral methods for sparse systems can be found in 
\cite{HuberSturm95,Jeronimo2008}.
Conditions on how far a Puiseux series should be expanded to decide
whether a point is isolated are given in~\cite{HJS15}.
The authors of~\cite{JLY16} propose numerical methods for tropical curves.
Polyhedral methods to compute tropical varieties are outlined
in~\cite{BJSST07} and implemented in Gfan~\cite{Jen08}.
The background on tropical algebraic geometry is in~\cite{MS15}.

Algorithms to compute the tropical {\em pre}variety 
are presented in~\cite{SV16}.  For {\em pre}processing purposes,
the software of~\cite{SV16} is useful.  However, the focus on this
paper concerns the tropical variety for which Gfan~\cite{Jen08}
provides a tropical basis.  Therefore, our computational experiments
with computer algebra methods are performed with Gfan and not with
the software of~\cite{SV16}.

\noindent {\bf Organization and Contributions.}
In the next section we illustrate the advantages of looking for
Puiseux series as solutions of polynomial systems.
Then we motivate our problem with some illustrative examples.
Relating the tropical prevariety to a recursive formula to compute
the mixed volume characterizes the generic case, in which the tropical
prevariety suffices to compute all space curve solutions.
With polyhedral end games we can recover the tropisms 
contained in higher dimensional cones of the tropical prevariety.
Finally we give some experimental results and timings.

\section{Puiseux Series}

\begin{figure}[hbt]
\begin{center}
\includegraphics[scale=.29]{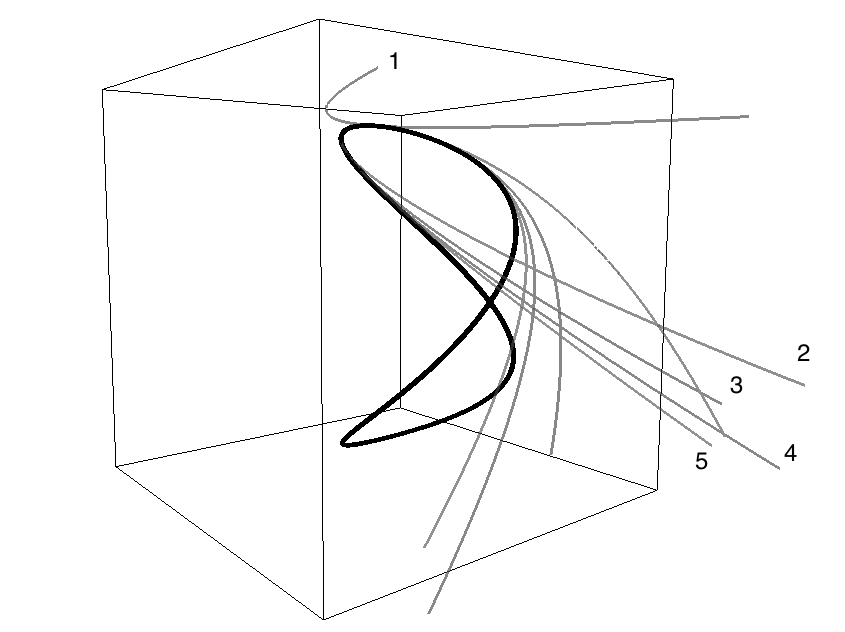}
\caption{Viviani's curve with improving Puiseux series 
approximations, labelled with the number of terms used to plot each one.}
\label{figviviani}
\end{center}
\end{figure}

When we work with Puiseux series we apply a hybrid method,
combining exact and approximate calculations.
Figure~\ref{figviviani} shows the plot, in black, of 
Viviani's curve, defined as the intersection of the sphere
$f = x_1^2 + x_2^2 + x_3^2 - 4 = 0$ 
and the cylinder $g = (x_1-1)^2 + x_2^2 -1= 0$. 

There is one pretropism $\bfv = (2,1,0)$,
which defines the initial forms of $f$ and $g$ respectively as
$x_3^2 - 4$ and $x_2^2 - 2x_1$.
For traditional Puiseux series, one would choose to set $x_1=1$,
obtaining the four solutions $(1,\pm \sqrt{2}, \pm 2)$ and leading
terms $(t^2, \pm \sqrt{2} t, \pm 2)$. If we instead use $x_1=2$,
we obtain rational coefficients and the following partial expansion:
\begin{equation}
\left[ \begin{array}{c} x_1 \\ x_2 \\ x_3 \end{array} \right]=\left[
\begin{array}{c} 2 t^{2} \\ 2t - t^{3}- \frac{1}{4}  t^{5}-\frac{1}{8}t^{7}
-\frac{5}{64}t^9\\
 2 - t^{2} - \frac{1}{4}t^{4} -\frac{1}{8}t^{6} -\frac{5}{64}t^9
\end{array}\right].
\end{equation}
The plot of several Puiseux approximations to Viviani's curve is shown in
gray in Figure~\ref{figviviani}.

If we shift the Viviani example so that its self-intersection is at
the origin, we obtain the following:
\begin{equation} \label{eqillusex}
   \bff(\x) = 
   \left\{
      \begin{array}{r}
        x_1^2 + x_2^2 + x_3^2 + 4x_1 =0 \\
        x_1^2 + x_2^2 + 2x_1 =0
      \end{array}
   \right.
\end{equation}
An examination of the first few terms of the Puiseux series expansion
for this system, combined with the On-Line Encyclopedia of Integer
Sequences~\cite{OEIS} and some straightforward algebraic manipulation,
allows us to hypothesize the following exact parameterization of the
variety:
\begin{equation}
\left[ \begin{array}{c} x_1 \\ x_2 \\ x_3 \end{array} \right]=\left[
\begin{array}{c} -2t^2 \\ 2\frac{t^3}{1 + \sqrt{1-t^{2}}}\; -\; 2t \\
-2t \end{array}\right].
\end{equation}
We can confirm that this is indeed right via substitution. While this
method is of course not possible in general, it does provide
an example of the potential usefulness of Puiseux series computations
for some examples.

\section{Assumptions and Setup}
Our object of study is space curves, by which we mean 1-dimensional
varieties in $\mathbb{C}^n$. Because Puiseux series computations take
one variable to be a free variable, we require that the curves not lie 
inside $V(\langle x_i\rangle)$ for some $i$; without loss of generality we choose
to use the first variable. Some
results require that the curve be in Noether position with respect to $x_1$,
meaning that the degree of the variety is preserved under intersection
with $x_1=\lambda$ for a generic $\lambda\in\mathbb{C}$. It is
of course possible to apply a random coordinate transformation to obtain
Noether position, but we then lose the sparsity of the system's exponent
support structure, which is what makes polyhedral methods effective.

\section{Some Motivating Examples}

In this section we illustrate the problem our paper addresses
with some simple examples, first in 3-space,
and then with a family of space curves in any dimensional space.

\subsection{In 3-Space}

Our first running example is the system
\begin{equation} \label{eqmotivex}
   \bff(\x) = 
   \left\{
      \begin{array}{r}
         x_1 x_3 - x_2 x_3 - x_3^2 + x_1 = 0 \\
         x_3^3 - x_1 x_2 - x_2 x_3 - x_3^2 - x_1 = 0
      \end{array}
   \right.
\end{equation}
which has an irreducible quartic and the second coordinate 
axis $(0,x_2,0)$ as its solutions.
Because the line lies in the first coordinate plane $x_1=0$,
the system is not in Noether position with respect to the first variable.
Therefore, our methods will ignore this part of the solution set.
The algorithms of~\cite{HJS12} can be applied to compute components
inside coordinate planes. Computing a primary decomposition yields the
following alternative, which lacks the portion in the first coordinate
plane:
\begin{equation} \label{eqmotivex2}
   \bf\tilde{f}(\x) = 
   \left\{
      \begin{array}{r}
        x_1 x_3 -  x_2 x_3 -  x_3^{2} + x_1 \\
        x_1 x_2 -  x_2^{2} -  x_2 x_3 + x_3^{2} + x_1 - 2 x_2 - 2 x_3 \\
         x_3^{3} -  x_2^{2} - 2 x_2 x_3 - 2 x_2 - 2 x_3
      \end{array}
   \right.
\end{equation}

The tropical prevariety contains the rays
$(2,1,1)$, $(1,0,0)$, and~$(1,0,1)$; because our Puiseux series start
their development at $x_1 = 0$,
rays that have a zero or negative value for their first coordinate 
have been discarded.  The tropical variety however contains the ray
$(3,1,1)$ instead of $(2,1,1)$, leading to the Puiseux expansion
\begin{equation}
\left[ \begin{array}{c} x_1 \\ x_2 \\ x_3 \end{array} \right]=\left[
    \begin{array}{c}
        108 t^3 \\
        t - 3 t^2 - 15 t^3 + 27 t^4 + 36 t^5 \\
        -t - 3 t^3 - 18 t^4 + 18 t^5 + 162 t^6
    \end{array}\right].
\end{equation}

This ray is a positive combination of $(2,1,1)$ and $(1,0,0)$.
In other words, it is possible for the 1-dimensional cones of the
tropical prevariety to fail to be in the tropical variety, and for rays
in the tropical variety to ``hide'' in the higher-dimensional cones of the
prevariety. 

\subsection{In Any Dimensional Space}

This problem can also occur in arbitrary dimensions, as seen in the
class of examples
\begin{equation} \label{eqgenexample}
   \bff(\x) = 
   \left\{
      \begin{array}{rcl}
        x_1^2 - x_1 + x_2 + x_3 + \cdots + x_n & = & 0 \\
        x_2^2 + x_1 + x_2 + x_3 + \cdots + x_n & = & 0 \\
        x_3^2 + x_1 + x_2 + x_3 + \cdots + x_n & = & 0 \\
                                            & \vdots & \\
        x_{n-1}^2 + x_1 + x_2 + x_3 + \cdots + x_n & = & 0.
      \end{array}
   \right.
\end{equation}
The ray $(1,1,1,\ldots,1)$ is a 1-dimensional cone of its prevariety since
it is normal to a facet of each polytope, namely the linear portion of
each polynomial. It is not, however, in the tropical variety, since the
initial form system (as it will be defined in Section~\ref{genericcase})
contains the monomial $x_1$.


\section{The Generic Case}\label{genericcase}
This hiding of tropisms in the higher dimensional cones of the
prevariety is problematic, as finding the
tropical variety may require more expensive symbolic computations.
For a comparison between various approaches see Section~\ref{compExp}.
Fortunately, this problem does not occur in general, as
the next result will show. But first, a few definitions.
\begin{definition} {\em
We write a polynomial $f$ with support set~$A$ as
\begin{equation}
   f(\x) = \sum_{\bfa \in A} c_\bfa \x^\bfa, \quad c_\bfa \in \cc^*,
   \x^\bfa = x_1^{a_1} x_2^{a_2} \cdots x_n^{a_n}.
\end{equation}
The {\em initial form of~$f$ with respect to $\bfv$} is then
\begin{equation}
   {\rm in}_\bfv f(\x) = \sum_{\bfa \in {\rm in}_\bfv A}
    c_\bfa \x^\bfa,
\end{equation}
where $\displaystyle {\rm in}_\bfv A = \{ \ \bfa \in A \ | \ 
  \langle \bfa, \bfv \rangle
  = \min_{\bfb \in A} \langle \bfb, \bfv \rangle \ \}$. }
\end{definition}
The initial form of a tuple of polynomials
is the tuple of the initial forms of the polynomials in the tuple.

\begin{definition} {\em
For $f\in\mathbb{C}[{\bf x}]$, $I$ an ideal in $\mathbb{C}[{\bf x}]$  and
${\bf v}\in \mathbb{R}^n$, we define the {\em initial ideal
$\text{in}_{\bf v}(I)$} as the ideal generated by $\{\text{in}_{\bf v}(f) : f\in I\}$. }
\end{definition}

\begin{definition} {\em
For $I=\langle f_1,\ldots,f_m \rangle \subset \mathbb{C}[{\bf x}]$ an ideal,
{\em the tropical prevariety}
is the set of ${\bf v}\in \mathbb{R}^n$ for which
$\text{in}_{\bf v}(f_i)$ is not a monomial for any $i$. The {\em tropical
variety} is the set of ${\bf v}\in \mathbb{R}^n$ for which
$\text{in}_{\bf v}(f)$ is not a monomial for any $f\in I$. }
\end{definition}

\begin{proposition}
For $n$ equations in $n+1$ unknowns with generic coefficients, the
set of ray generators of the tropical prevariety contains the tropical variety.
\end{proposition}

It is important to note that our notion of generic here refers to the
coefficients, and not to generic tropical varieties as seen in
\cite{Rom12} which are tropical varieties of ideals under a generic
linear transformation of coordinates.

The tropical prevariety always contains the tropical variety. We simply
want to show that all of the rays of the tropical variety show up in the
prevariety as ray generators, and not as members of the
higher-dimensional cones. Let
$I=\langle p_1,\ldots,p_n\rangle \subseteq \mathbb{C}[x_0,\ldots,x_n]$,
and let $\mathbf{w}$ be a ray in the tropical prevariety but not one of its
ray generators. We want to show that $\mathbf{w}$ is not in the tropical
variety, or equivalently that $\text{in}_\mathbf{w}(I)$ contains a monomial.
We will do so by showing that $I_\mathbf{w}:=\langle
\text{in}_\mathbf{w}(p_1),\ldots,\text{in}_\mathbf{w}(p_n)\rangle$
contains a monomial, which suffices since this ideal is contained in
$\text{in}_\mathbf{w}(I)$.

Suppose $I_\mathbf{w}$ contains no monomial. Then
$(x_0x_1\cdots x_n)^k\notin I_\mathbf{w}$ for any $k$. By Hilbert's 
Nullstellensatz $V:=\mathbb{V}(I_\mathbf{w})\nsubseteq \mathbb{V}(x_0x_1\cdots x_n)$,
i.e.\ $V$ is not contained in the union of the coordinate hyperplanes.
Then there exists $a=(a_0,\ldots,a_n)\in V$ such that all coordinates
of $a$ are all nonzero. Since $\mathbf{w}$ lies in the interior a cone
of dimension at least 2, the generators of $I_\mathbf{w}$ are homogeneous
with respect to at least two linearly independent rays {\bf u} and {\bf v}.
Thus
$(\lambda^{\mathbf{u}_0}\mu^{\mathbf{v}_0} a_0,\ldots,\lambda^{\mathbf{u}_n}
\mu^{\mathbf{v}_n} a_n)\in V$ for all $\lambda,\mu\in\mathbb{C}\setminus\{0\}$
where the $\mathbf{u}_i,\mathbf{v}_i$ are the components of $\bf u$
and $\bf v$, and $V$ contains a toric surface. If we intersect with a random
hyperplane, by Bernstein's theorem B \cite{Ber75} the result is a finite 
set of points, with the possibility of additional components that must
be contained in the coordinate planes. Hence $V$ can contain no surface
outside of the coordinate planes, and we have a contradiction.

\hfill~\qed

\section{Polyhedral Methods}

We will show that the tropical prevariety provides an upper bound
for the degree of the solution curve.
The inner product of a point $\bfa$ with a vector $\bfv$ is denoted 
as $\langle \bfa , \bfv \rangle = a_1 v_1 + a_2 v_2 + \cdots + a_n v_n$.

\begin{lemma} \label{lemvol}
Consider an $(n-1)$-tuple of Newton polytopes
${\cal P} = (P_1, P_2, \ldots, P_{n-1})$ in $n$-space.
Let $E$ be the edge spanned by $(1,0,\ldots,0)$
and $(0, 0, \ldots, 0)$.
The mixed volume of $({\cal P}, E)$ equals
\begin{equation}
   V_n({\cal P}, E)
   = \sum_{\bfv} v_1 V_{n-1}({\rm in}_\bfv {\cal P}),
\end{equation}
where $\bfv$ ranges over all rays in the tropical prevariety
of $\cal P$ with $v_1 > 0$, normalized so that 
${\rm gcd}(\bfv) = {\rm gcd}(v_1,v_2, \ldots, v_n) = 1$,
and ${\rm in}_\bfv {\cal P}
= ({\rm in}_\bfv P_1, {\rm in}_\bfv P_2, \ldots$, 
  ${\rm in}_\bfv P_{n-1})$, where
${\rm in}_\bfv P_k$ is the face with support vector $\bfv$,
formally expressed as
\begin{equation}
   {\rm in}_\bfv P_k
    = \{ \ \bfa \in P_k \ | \ \langle \bfa , \bfv \rangle
    = \max_{\bfa \in P_k} \langle \bfa, \bfv \rangle \ \}.
\end{equation}
\end{lemma}
We apply the following recursive formula~\cite{Sch93}
for the mixed volume 
\begin{equation} \label{eqmvformula}
   V_n({\cal P}, E) =
   \sum_{ \begin{array}{c}
             \bfv \in \zz^n \\ \gcd(\bfv) = 1
          \end{array} } \ p_E (\bfv) \ V_{n-1}({\rm in}_\bfv {\cal P}),
\end{equation}
where $p_E$ is the support function of the edge~$E$:
\begin{equation}
   p_E(\bfv) = \max_{\bfe \in E} \langle \bfe, \bfv \rangle
\end{equation}
and ${\rm in}_\bfv {\cal P}
= ({\rm in}_\bfv P_1, {\rm in}_\bfv P_2, \ldots, 
   {\rm in}_\bfv P_{n-1})$, where
\begin{equation}
   {\rm in}_\bfv P_k
    = \{ \ \bfa \in P_k \ | \ \langle \bfa , \bfv \rangle = p_k(\bfv) \ \},
\end{equation}
with $p_k$ the support function of the polytope $P_k$.

Because the edge $E$ contains $(0,0,\ldots,0)$: $p_E(\bfv) \geq 0$
and $p_E(\bfv) = 0$ when $v_1 \leq 0$.  Only those rays for which
$v_1 > 0$ contribute to $V_n({\cal P}, E)$.
We have then $p_E(\bfv) = v_1$.

The mixed volume of a tuple of polytopes equals zero
if one of the polytopes consists of only one vertex.
The rays in the tropical prevariety contain all vectors for which
${\rm in}_\bfv(P_k)$ is an edge or a higher dimensional face.
These are the rays $\bfv$ for which 
$V_{n-1}({\rm in}_\bfv {\cal P}) > 0$.

\hfill~\qed

The application of Lemma~\ref{lemvol} leads to a bound on 
the number of generic points on the space curve.
Denote $\cc^* = \cc \setminus \{ 0 \}$.

\begin{lemma}
Consider the system $\bff(\x) = \zero$,
$\bff = (f_1$, $f_2$, $\ldots$, $f_{n-1})$
with ${\cal P} = (P_1, P_2, \ldots, P_{n-1})$
where $P_k$ is the Newton polytopes of~$f_k$.
If the system is in Noether position with respect to $x_1$,
then the degree of the space curve defined by $\bff(\x) = \zero$
is bounded by $V_n({\cal P}, E)$.
\end{lemma}

This result is a version of Lemma 2.3 from \cite{Jeronimo2008}.

The proof of the lemma follows from the application
of Bernshtein's theorem~\cite{Ber75} to the system
\begin{equation}
  \left\{
     \begin{array}{l}
       \bff(\x) = \zero \\
       x_1 = \gamma, \quad \gamma \in \cc^*.
     \end{array}
  \right.
\end{equation}
By the assumption of Noether position, 
there will be as many solutions to this system
as the degree of the space curve defined by~$\bff(\x) = \zero$.
The theorem of Bernshtein states that the mixed volume bounds
the number of solutions in $(\cc^*)^n$.

\hfill~\qed

Formula~(\ref{eqmvformula}) appears in the constructive proof
of Bernshtein's theorem~\cite{Ber75} and was implemented
in the polyhedral homotopies of~\cite{VVC94}.
For systems with coefficients that are sufficiently generic,
the mixed volumes provide an exact root count.

\begin{theorem}
Let $\bff(\x) = \zero$ be a polynomial system of $n-1$ equations
in $n$ unknowns, with sufficiently generic coefficients.
Assume the space curve defined by $f(\x) = \zero$ is in Noether
position with respect to the first variable.
Then all rays $\bfv$ with $v_1 > 0$ in the tropical prevariety of $\bff$
lead to Puiseux series expansions for the space curve defined by
$\bff(\x) = \zero$.  Moreover, the degree of the space curve is the
sum of the degrees of the Puiseux series.
\end{theorem}

We illustrate the application of polyhedral methods
to the motivating examples.

\begin{example} {\rm
As a verification on the first motivating example~(\ref{eqmotivex}),
we consider the rays $(2,1,1)$, $(1,0,0)$, and~$(1,0,1)$
of its tropical prevariety. 
The initial form of $\bff$ in~(\ref{eqmotivex}) w.r.t. to
the ray $(2,1,1)$ is
\begin{equation}
   {\rm in}_{(2,1,1)} \bff(\x) =
   \left\{
      \begin{array}{l}
         -x_2 x_3 - x_3^2 + x_1 = 0 \\
         -x_2 x_3 - x_3^2 - x_1 = 0. \\
      \end{array}
   \right.
\end{equation}
To count the number of solutions of ${\rm in}_{(2,1,1)} \bff(\x) = \zero$
we apply a unimodular coordinate transformation, $\x = \y^U$:
\begin{equation}
   U =
   \left[
     \begin{array}{ccc}
        2 & 1 & 1 \\
        1 & 0 & 0 \\
        0 & 1 & 0 \\
     \end{array}
   \right]
   \quad
    \left\{
       \begin{array}{lclll}
          x_1 & = & y_1^2 & y_2 & \\
          x_2 & = & y_1   &     & y_3 \\
          x_3 & = & y_1   &     &  \\
       \end{array}
    \right.
\end{equation}
which leads to the system
\begin{equation}
   {\rm in}_{(2,1,1)} \bff(\y) =
   \left\{
      \begin{array}{l}
         -y_1^2 y_3 - y_1^2  + y_1^2 y_2 = 0 \\
         -y_1^2 y_3 - y_1^2  - y_1^2 y_2 = 0. \\
      \end{array}
   \right.
\end{equation}
After removing the common factor $y_1^2$, we see that this system
has one solution for generic choices of the coefficients.
As 2 is the first coordinate of $(2,1,1)$, this ray contributes
two branches and adds two to the degree of the solution curve.
The other rays $(1,0,0)$ and $(1,0,1)$ each contribute one to 
the degree, and so we recover the degree four of the solution curve. }
\end{example}

\begin{example} {\rm
For the family of systems in~(\ref{eqgenexample}), consider
the curve in 4-space:
\begin{equation}
   \bff(\x) = 
   \left\{
      \begin{array}{l}
          x_1^2 - x_1 + x_2 + x_3 + x_4 = 0 \\
          x_2^2 + x_1 + x_2 + x_3 + x_4 = 0 \\
          x_3^2 + x_1 + x_2 + x_3 + x_4 = 0.
      \end{array}
   \right.
\end{equation}
For the tropism $\bfv = (2,1,1,1)$, the initial form is
\begin{equation}
   {\rm in}_\bfv \bff(\x) = 
   \left\{
      \begin{array}{l}
          x_2 + x_3 + x_4 = 0 \\
          x_2 + x_3 + x_4 = 0 \\
          x_2 + x_3 + x_4 = 0.
      \end{array}
   \right.
\end{equation}
This tropism is in the interior of the cone in the tropical prevariety
spanned by $v_1=(1,1,1,1)$ and $v_2=(1,0,0,0)$. Using the same
techniques as in the previous example, we find  $\text{in}_{\bf v_1}(I)$
has a mixed volume of one and $\text{in}_{\bf v_2}(I)$ has a mixed
volume of three, so for generic coefficients we again recover the degree
of the solution curve.
}
\end{example}

\section{Current Approaches}\label{currentApproaches}
In \cite{BJSST07} a method is given for computing the tropical variety of
an ideal $I$ defining a curve. It involves appending witness polynomials
from $I$ to a list of its generators such that for this new set, the
tropical prevariety equals the tropical variety. Such a set is called a
{\em tropical basis}. Each additional polynomial rules out one of the
cones in the original prevariety that does not belong in the tropical
variety. As stated in \cite{BJSST07} only finitely many additional
polynomials are necessary, since the prevariety has only finitely many
cones.

The algorithm runs as follows. For each cone $C$ in the tropical
prevariety, we choose a generic element ${\bf w}\in C$. We check whether
in$_{\bf w}(I)$ contains a monomial by saturating with respect to $m$,
the product of ring variables; the initial ideal contains a monomial if
and only if this saturation ideal is equal to $(1)$. If in$_{\bf w}(I)$
does not contain a monomial, the cone $C$ belongs in
our tropical variety. If it does, we check whether $m^i\in I$ for
increasing values of $i$ until we find a monomial
$m'\in \text{in}_{\bf w}(I)$. Finally, we append $m'-h$ to our list of
basis elements, where $h$ is the reduction of $m$ with respect to a
Gr{\"o}bner basis of $I$ under any monomial order that refines ${\bf w}$.
For ${\bf w}$ to define a global monomial order, and thus allow a
Gr\"obner basis, it may be necessary to homogenize the ideal first.

Bounding the complexity of this algorithm is beyond the scope of this
paper, but for each cone it requires computing a Gr\"obner basis of $I$
as well as another (possibly faster) basis when calculating the
saturation to check if the initial ideal contains a monomial. In some
cases we may only be concerned about tropisms hiding in a
particular higher-dimensional cone of the prevariety, such as with
our running example (\ref{eqgenexample}). Here it is reasonable to perform
only one step of this algorithm, namely looking for a witness for a single
cone, which could be significantly faster. However, this has the
disadvantage of introducing more 1-dimensional cones into the prevariety.
More details, including some timing comparisons, will be given in
Section~\ref{compExp}.

\section{Polyhedral End Games}

A polyhedral end game~\cite{HV98} applies extrapolation methods
to numerically estimate the winding number of solution paths
defined by a homotopy.  The leading exponents of the Puiseux series
are recovered via differences of the logarithms of the magnitudes
of the coordinates of the solution paths.
Even in the case -- as in our illustrative example -- where
the given polynomials contain insufficient information to compute
all tropisms only from the prevariety, a polyhedral end game 
manages to compute all tropisms.  
The setup is similar to that of~\cite{Ver09}, arising in a numerical 
study of the asymptotics of a space curve, 
defined by the system $\bff(\x) = \zero$:
\begin{equation} \label{eqhomotopy}
   \left\{
      \begin{array}{c}
         \bff(\x) = \zero \\
         t x_1 + (1-t) (x_1 - \gamma) = 0,
         \quad \gamma \in \cc \setminus \{ 0  \},
      \end{array}
   \right.
\end{equation}
as $t$ moves from 0 to 1, the hyperplane $x_1 = \gamma$
moves the coordinate plane perpendicular to the first coordinate axis.

As $t$ moves from 0 to 1, it is important to note that
$t$ will actually never be equal to one.
In the polyhedral end games of~\cite{HV98},
to estimate the winding number via extrapolation methods,
the step size decreases in a geometric ratio.
In particular, denoting the winding number by $\omega$,
for $t = 1 - s^\omega$, and $0 < r < 1$,
we consider the solutions for $s_k = s_0 r^k$, $k = 0, 1, \ldots$,
starting at some $s_0 \approx 0$.

The constant $\gamma$ in~(\ref{eqhomotopy}) is a randomly generated
complex number.  This implies that for $x_1 = \gamma$, 
the polynomial system in~(\ref{eqhomotopy}) for $t=0$
has as many isolated solutions (generic points on the space,
eventually counted with multiplicities) as the degree of the projection
of the space curve onto the first coordinate plane.
As long as $t < 1$, the points remain generic, although the numerical
condition numbers are expected to blow up as $t$ approaches one.

The deteriorating numerical ill conditioning can be mitigated
by the use of multiprecision arithmetic.
For example, condition numbers larger than $10^8$ make results 
unreliable in double precision.  In double double precision,
much higher condition numbers can be tolerated, typically up to~$10^{16}$,
and this goes up to~$10^{32}$ for quad double precision.
As we interpret the inverse of the condition number as the distance
to a singular solution, with multiprecision arithmetic we can compute
more points more accurately as needed in the extrapolation to estimate
winding numbers.

An additional difficulty arises when a path diverges to infinity,
which manifests itself by a tropism with negative coordinates.
A reformation of the problem in a weighted projective space
corresponds to a unimodular coordinate transformation which uses
the computed direction of the solution path.  Towards the end of
the path, this direction coincides with the tropism.

The a posteriori verification of a polyhedral end game
is similar to computing a Puiseux expansion starting at a pretropism.

\section{Computational Experiments}\label{compExp}

In this section we focus on the family of systems~(\ref{eqgenexample})
with a tropism hidden in a higher dimensional cone of pretropisms.
Classical families such as the cyclic $n$-roots problems appear
not to have such hidden pretropisms, at least not for the cases
computed in~\cite{AV12,AV13} and~\cite{Sab11}.

\subsection{Symbolic Methods}

To substantiate the claim that finding the tropical variety is
computationally expensive, we calculated tropical bases of the
system~(\ref{eqgenexample}) for various values of $n$.
The symbolic computations of tropical bases was done
with Gfan~\cite{Jen08}.   
Times are displayed in Figure~\ref{tropbasistimes}.
The computations were executed on an Intel Xeon E5-2670 processor
running RedHat Linux. As is clear from the table, as the dimension grows for
this relatively simple system, computation time becomes prohibitively large.
\begin{table}[hbt]
\centering
\caption{Execution times, in seconds, of the computation of a tropical
basis for the system~(\ref{eqgenexample}); averages of 3 trials.}
\def\arraystretch{1.1}%
\begin{tabular}{c|ccccc}
\noalign{\smallskip}
n    & 3     &     4 &     5 &      6 &     7 \\ \hline
time & 0.052 & 0.306 & 2.320 & 33.918 & 970.331 \\
\noalign{\smallskip}
\end{tabular}
\label{tropbasistimes}
\end{table}

As mentioned in Section~\ref{currentApproaches}, an alternative to
computing the tropical basis is to only calculate the witness polynomial
for a particular cone of the tropical prevariety. We implemented this
algorithm in Macaulay2 \cite{M2} and applied it to~(\ref{eqgenexample}) to
cut down the cone generated by the rays $(1,1,\ldots,1)$ and
$(1,0,0,\ldots,0)$. In all the cases we tried, the new prevariety contained
the ray $(2,1,\ldots,1)$, as we expected.

From Table~\ref{cutCone} it is clear that this has a significant speed
advantage over computing a full tropical basis. However, it has the
disadvantage of introducing many more rays into the prevariety. The number
can vary depending on the random ray chosen in the cone, so the third column
lists some of the values we obtained over several trials. We only computed up
through dimension 10 because the prevariety computations were excessive for
higher dimensions.

\begin{table}[hbt]
\centering
\caption{Execution times in seconds
of the computation of a witness polynomial for
the cone generated by $(1,1,\ldots,1),(1,0,\ldots,0)$ of the
system~(\ref{eqgenexample}); averages of 3 trials. The third column lists
the number of rays in the fan obtained by intersecting the original prevariety
with the normal fan of the witness polynomial; since this can vary with the choice
of random ray, we list values from several tries.}
\bgroup
\def\arraystretch{1.1}%
\begin{tabular}{c|r|l}
dim & time~~  & \#rays in the new fan\\ \hline
 3     &  0.004    &  4, 5        \\
 4     &  0.011    &  10, 11       \\
 5     &  0.004    &  13, 14       \\
 6     &  0.009    &  27, 49       \\
 7     &  0.033    &  13, 25, 102    \\
 8     &  0.170    &  124, 401, 504    \\
 9     &  0.963    &  758, 1076 \\
 10    &  10.749   &  514, 760, 1183, 2501     \\
 11    &  131.771  &       \\
 12    &  1131.089 &       \\
\end{tabular}
\egroup
\label{cutCone}
\end{table}

\subsection{Our Approach}

The polyhedral end game was done with version 2.4.10 of PHCpack~\cite{Ver99},
upgraded with double double and quad double arithmetic,
using QDlib~\cite{HLB00}.  Polyhedral end games are also available via
the Python interface of PHCpack, since version 0.4.0 of phcpy~\cite{Ver14}.

For the first motivating example~(\ref{eqmotivex}) in 3-space,
there are four solutions when $x_1 = \gamma$.
The tropism $(3,1,1)$, with winding number 3, is recovered
when running a polyhedral end game, tracking four solution paths.
Even in quad double precision (double precision already suffices),
the running time is a couple of hundred milliseconds.

Table~\ref{tabmotivfam} shows execution times for the 
family of polynomial systems in~(\ref{eqgenexample}).
The computations were executed on one core of
an Intel Xeon E5-2670 processor, running RedHat Linux.

\begin{table}[hbt]
\centering
\caption{Execution times on tracking $d$ paths in $n$-space with
a polyhedral end game.  The reported time is the elapsed CPU user time,
in seconds.  The last column represents the average time spent on one path.}
\begin{tabular}{r|r|r|r}
$n$ &  $d$ &   time  & time/d \\ \hline
  4 &    4 &   0.012 & 0.003 \\
  5 &    8 &   0.035 & 0.006 \\
  6 &   16 &   0.090 & 0.007 \\
  7 &   32 &   0.243 & 0.010 \\
  8 &   64 &   0.647 & 0.013 \\
  9 &  128 &   1.683 & 0.016 \\
 10 &  256 &   4.301 & 0.017 \\
 11 &  512 &   7.507 & 0.015 \\
 12 & 1024 &  27.413 & 0.027  \\
\end{tabular}
\label{tabmotivfam}
\end{table}

All directions computed with double precision
at an accuracy of $10^{-8}$.
Double precision sufficed to accurately
compute the tropism $(2,1,\ldots,1)$.
Although the total number of paths grows exponentially, every path
has the same direction, so tracking only one path suffices.
These times are significantly smaller than the time required to
compute a tropical basis.

\section{Conclusions}

The tropical prevariety provides candidate tropisms for Puiseux
series expansions of space curves.
As shown in~\cite{AV12,AV13} on the cyclic $n$-root problems,
the pretropisms may directly lead to series developments for 
the positive dimensional solution sets.
In this paper we studied cases where tropisms are in the relative
interior of higher-dimensional cones of the tropical prevariety.
If the tropical prevariety contains a higher dimensional cone
and Puiseux series expansion fails at one of the cone's generating
rays, then a polyhedral end game can recover the tropisms in the
interior of that higher dimensional cone of pretropisms. As our
example shows, this takes drastically less time than computing the
tropical variety via a tropical basis, especially as dimension grows.
It is also faster than finding a witness polynomial for just that
particular cone, and avoids the issue of adding rays to the tropical
prevariety.

\bibliographystyle{plain}

\end{document}